\documentclass[12pt]{article}
\usepackage{graphicx}
\usepackage{color}
\usepackage{cite}
\usepackage{here}

\makeatletter
\g@addto@macro\bfseries{\boldmath}
\makeatother

\def \beq{\begin{equation}}
\def \eeq{\end{equation}}
\def\eqref#1{(\ref{#1})}
\def\bea{\begin{eqnarray}}
\def\eea{\end{eqnarray}}
\def\jpsi{J\kern-0.1em/\kern-0.1em\psi}
\def\nl{\hfill\break}

\def\URLtilde{\lower0.2em\hbox{$\tilde{\phantom{a}}$}}
\def\mycomm#1{\hfill\break\strut\kern-3em{\color{red}\tt ====> #1
\color{black}}\hfill\break}

\newcount\timecount
\newcount\hours \newcount\minutes  \newcount\temp \newcount\pmhours
\hours = \time
\divide\hours by 60
\temp = \hours
\multiply\temp by 60
\minutes = \time
\advance\minutes by -\temp
\def\hour{\the\hours}
\def\minute{\ifnum\minutes<10 0\the\minutes
\else\the\minutes\fi}
\def\clock{
\ifnum\hours=0 12:\minute\ AM
\else\ifnum\hours<12 \hour:\minute\ AM
\else\ifnum\hours=12 12:\minute\ PM
\else\ifnum\hours>12
\pmhours=\hours
\advance\pmhours by -12
\the\pmhours:\minute\ PM
\fi
\fi
\fi
\fi
}

\def\monthname{\relax\ifcase\month 0/\or January\or February\or
March\or April\or May\or June\or July\or August\or September\or
October\or November\or December\else\number\month/\fi}

\def\bold#1{\setbox0=\hbox{$#1$}     \kern-.025em\copy0\kern-\wd0
\kern.05em\copy0\kern-\wd0
\kern-.025em\raise.0433em\box0 }

\begin{document}
\setcounter{footnote}{1}
\rightline{EFI 17-17}
\rightline{TAUP 3021/17}
\rightline{arXiv:1707.07666}
\vskip1.5cm

\begin{center}
{\large \bf \boldmath
Discovery of doubly-charmed $\Xi_{cc}$ baryon implies 
\\ \vrule width0pt height3ex
a stable $bb\bar u\bar d$ tetraquark
\unboldmath}
\end{center}
\bigskip

\centerline{Marek Karliner$^a$\footnote{{\tt marek@proton.tau.ac.il}}
 and Jonathan L. Rosner$^b$\footnote{{\tt rosner@hep.uchicago.edu}}}
\medskip

\centerline{$^a$ {\it School of Physics and Astronomy}}
\centerline{\it Raymond and Beverly Sackler Faculty of Exact Sciences}
\centerline{\it Tel Aviv University, Tel Aviv 69978, Israel}
\medskip

\centerline{$^b$ {\it Enrico Fermi Institute and Department of Physics}}
\centerline{\it University of Chicago, 5620 S. Ellis Avenue, Chicago, IL
60637, USA}
\bigskip
\strut

\begin{center}
ABSTRACT
\end{center}
\begin{quote}
Recently LHCb discovered the first doubly-charmed baryon $\Xi_{cc}^{++} =
ccu$ at $3621.40 \pm 0.78$ MeV, very close to our theoretical prediction.
We use the same methods to predict a doubly-bottom tetraquark \,$T(bb\bar
u\bar d)$\, with $J^P{=}1^+$ at $10,389\pm 12$ MeV,\, 215 MeV below the
$B^-\bar B^{*0}$ threshold and 170 MeV below threshold for decay to 
$B^-\bar B^0 \gamma$.  The
$T(bb\bar u\bar d)$\, is therefore stable under strong and electromagnetic
(EM) interactions and can only decay weakly, the first exotic hadron
with such a property.  On the other hand, the mass of \,$T(cc\bar u\bar
d)$\, with $J^P{=}1^+$ is predicted to be $3882\pm12$ MeV, 7 MeV above
the $D^0 D^{*+}$ threshold and 148 MeV above $D^0 D^+ \gamma$ threshold.
$T(bc\bar u\bar d)$ with $J^P{=}0^+$ is predicted at $7134\pm13$ MeV, 11 MeV
below the $\bar B^0 D^0$ threshold.  Our precision is not sufficient to
determine whether $bc\bar u\bar d$\, is actually above or below the threshold.
It could manifest itself as a narrow resonance just at threshold.
\end{quote}
\smallskip

\leftline{PACS codes: 14.20.Lq, 14.20.Mr, 12.40.Yx}
\bigskip


\section{INTRODUCTION \label{sec:intro}}

The question whether $QQ\bar q\bar q$ tetraquarks with two heavy quarks 
$Q$ and two light antiquarks $\bar q$ are stable or unstable against decay
into two $Q\bar q$ mesons has a long history. It has been largely
undecided, mainly due to lack of experimental information about
the strength of the interaction between two heavy quarks.

The very recent discovery of the doubly charmed baryon 
$\Xi_{cc}$ by the LHCb Collaboration at CERN has now provided the crucial 
experimental input which allows this issue to be finally resolved.

LHCb has observed the doubly-charmed baryon 
\,$\Xi_{cc}^{++} = ccu\,$ with a mass of $3621.40 \pm 0.78$ MeV 
\cite{Aaij:2017ueg}.  This value is consistent with several predictions,
including our value of $3627 \pm 12$ MeV \cite{Karliner:2014gca,note1}.

Here we use similar methods to those in Ref.\ \cite{Karliner:2014gca}
and earlier works \cite{Karliner:2008sv} to predict the mass of the
ground-state $bb\bar u\bar d$ tetraquark with spin-parity $J^P = 1^+$,
$M(T(bb\bar u\bar d)) = 10,389\pm 12$ MeV.

Angular momentum and parity conservation in strong and EM interactions forbid a
state with $J^P = 1^+$ from decaying strongly or electromagnetically into two
pseudoscalars in any partial wave. Therefore $bb\bar u\bar d$ with $J^P=1^+$
cannot decay into $BB$. The lowest-mass hadronic channel allowed by angular
momentum and parity is $B B^*$, most favorably in $S$-wave. This channel is
however kinematically closed, because the $T(bb\bar u\bar d)$ mass is 215 MeV
below $BB^*$ threshold at 10,604 MeV. $M(T(bb\bar u\bar d))$ is also 170
MeV below $2 m_B$, the relevant threshold for EM decay to $B^-\bar B^0 \gamma$.

The $B$ mesons are the lightest states that carry open bottom, so the 
$bb\bar u\bar d$ tetraquark cannot decay through strong or EM interactions 
which conserve heavy flavor. It can only
decay weakly, when one of the $b$ quarks decays into and $c$ quark
and a virtual $W^+$.  A typical decay is therefore $(bb\bar u\bar d)
\to \bar B D \pi^+(\rho^+)$, etc.

The main challenge in the prediction of the $T(bb\bar u\bar d)$ mass
is the estimate of binding energy between the two $b$ quarks 
\cite{Karliner:2006hf,Karliner:2013dqa,Luo:2017eub}.
Table IX of Ref.~\cite{Luo:2017eub} provides a compilation of earlier
mass estimates of various $QQ\bar q\bar q$ tetraquarks.
In Ref.\ \cite{Karliner:2014gca} we estimated the binding energy
between two heavy quarks $Q$ by assuming that $QQ$ binding is one-half of the
$\bar Q Q$ binding which can be obtained from quarkonia.
When applied to the $ccu$ baryon $\Xi_{cc}$ this led to the
prediction $M(\Xi_{cc}) =3627 \pm 12$ MeV, very close to
the experimentally measured $ccu$ mass of $3621.40 \pm 0.78$ MeV.

The above relation between quark-quark and quark-antiquark binding is exact
in the one-gluon-exchange weak-coupling limit. Its successful 
extension beyond weak coupling implies that the heavy quark 
potential factorizes into a color-dependent and a space-dependent part,
with the space-dependent part being the same for $QQ$ and $\bar Q Q$.
The relative factor 1/2 is then automatic, just as in the weak coupling
limit, resulting from the color algebra.

\section{CALCULATION OF THE $bb\bar u\bar d$ MASS \label{sec:bbmass}}
In the present work we build on the accuracy of the $\Xi_{cc}$ mass
prediction and assume the same relation is true for $bb$ binding energy
in a $bb\bar u\bar d$ tetraquark.  In order to obtain a state with the
lowest possible mass, we further assume that all four quarks are in a
relative $S$-wave and that the $\bar u$ and $\bar d$ light antiquarks
bind into a color-triplet ``good" antidiquark with spin and isospin zero.
The $bb$ diquark must then be a color antitriplet and Fermi statistics
dictates it has spin 1. The total spin and parity are then $J^P = 1^+$.

The upshot is that we are considering a configuration very similar to
a heavy-light meson $\bar Q q$, where instead of the heavy antiquark
we have a doubly-heavy color antitriplet diquark and instead of the 
quark we have a light color triplet antidiquark.
The rest of the calculation is straightforward and proceeds in a way
entirely analogous to Ref.~\cite{Karliner:2014gca}.

The contributions to the mass of the lightest tetraquark $T(bb\bar
u\bar d)$ with two bottom quarks and $J^P=1^+$ are listed in
Table \ref{tab:bbud}.  The notation and the numerical values of
all the parameters are the same as in Table VI and Table IX of
Ref.~\cite{Karliner:2014gca}.  In particular, the subscripts on masses $m$
denote flavor, while the superscripts $b$ indicate that these are effective
masses in baryons.

The central value of the resulting mass \,10,389 MeV$\pm12$ \,is\, 215 MeV below
$BB^*$ threshold at 10,604 MeV, and 170 MeV below $B^-\bar B^0\gamma$ 
threshold at 10,559 MeV.

\begin{table}
\caption{Contributions to the mass of the lightest tetraquark 
$T(bb\bar u\bar d)$ with two bottom quarks and $J^P=1^+$.
\label{tab:bbud}}
\begin{center}
\begin{tabular}{c r} \hline \hline
Contribution & Value (MeV) \\ \hline
$2m^b_b$ & 10087.0\\
$2m^b_q$ & 726.0\\
$a_{bb}/(m^b_b)^2$ & 7.8 \\
${-}3a/(m^b_q)^2$ & ${-}150.0$ \\
$bb$ binding & ${-}281.4$ \\
Total & $10389.4 \pm 12$ \kern-2.55em \\ \hline \hline
\end{tabular}
\end{center}
\end{table}

\section{$cc\bar u\bar d$ \,and\, $bc\bar u\bar d$\, MASSES \label{sec:ccmass}}
The calculation of the masses of the lightest \,$cc\bar u\bar d$\, and\,
tetraquark masses proceeds analogously to $bb\bar u\bar d$.
In Tables~\ref{tab:ccud} and~\ref{tab:bcud} we provide the corresponding
contributions to the $cc\bar u\bar d$ and $bc\bar u\bar d$ masses.

The mass of $cc\bar u\bar d$ turns out to be $3882\pm12$ MeV, with the central
value only 7 MeV above the \,$D^0 D^{*+}$ \,threshold at 3875 MeV
and 148 MeV above $D^0 D^+ \gamma$ threshold.
Moreover, as the central value of our prediction of $M(\Xi_{cc}^{++})$ is 6
MeV above the observed central value, if we were to increase the $cc$ binding
energy by 6 MeV to force agreement between prediction and observation, the
mass $cc\bar u\bar d$ would be lowered to 3876 MeV, only 1 MeV above 
\,$D^0 D^{*+}$ \, threshold.  As $M(D^0) + M(D^{*+}) = 3875.09 \pm 0.07$ MeV
while $M(D^+) + M(D^{*0})$ is $1.35\pm0.12$ MeV higher at $3876.44 \pm 0.10$
MeV \cite{PDG}, there may be some interesting violations of isospin in the
hadronic decays of such a state, in analogy with isospin violations in decays
of $X(3872)$ \cite{Tornqvist:2004qy}.

Unlike $bb\bar u\bar d$ and $cc\bar u\bar d$,
the lowest mass $bc\bar u\bar d$ tetraquark has $J^P=0^+$,
because the minimal energy $bc$ diquark has spin zero.
The $bc\bar u\bar d$ mass is $7133.7\pm13$ MeV, with the central value 
about 11 MeV {\em below}\, the \,$\bar B^0 D^0$ \,threshold at 7144.5 MeV.

The precision of our calculation is not sufficient to determine whether the
$bc\bar u\bar d$ tetraquark is actually above or below the corresponding
two-meson threshold.  It could manifest itself as a narrow resonance just at
threshold.

\begin{table}[H]
\caption{Contributions to the mass of the lightest tetraquark 
$T(cc\bar u\bar d)$ with two charmed quarks and $J^P=1^+$.
\label{tab:ccud}}
\begin{center}
\begin{tabular}{c r} \hline \hline
Contribution & Value (MeV) \\ \hline
$2m^b_c$ & 3421.0\\
$2m^b_q$ & 726.0\\
$a_{cc}/(m^b_c)^2$ & 14.2 \\
${-}3a/(m^b_q)^2$ & ${-}150.0$ \\
$cc$ binding & ${-}129.0$ \\
Total & $3882.2 \pm 12$ \kern-2.55em \\ \hline \hline
\end{tabular}
\end{center}
\end{table}

\begin{table}[H]
\caption{Contributions to the mass of the lightest tetraquark 
$T(bc\bar u\bar d)$ with one bottom and one charmed quark and $J^P=0^+$.
\label{tab:bcud}}
\begin{center}
\begin{tabular}{c r} \hline \hline
Contribution & Value (MeV) \\ \hline
$m_b^b + m^b_c$ & 6754.0\\
$2m^b_q$ & 726.0\\
${-}3a_{bc}/(m_b^b m^b_c)$ & ${-}$25.5 \\
${-}3a/(m^b_q)^2$ & ${-}150.0$ \\
$bc$ binding & ${-}170.8$ \\
Total & $7133.7\pm 13$ \kern-2.55em \\ \hline \hline
\end{tabular}
\end{center}
\end{table}

\begin{figure}[t]
\begin{center}
\includegraphics[width=0.95\textwidth]{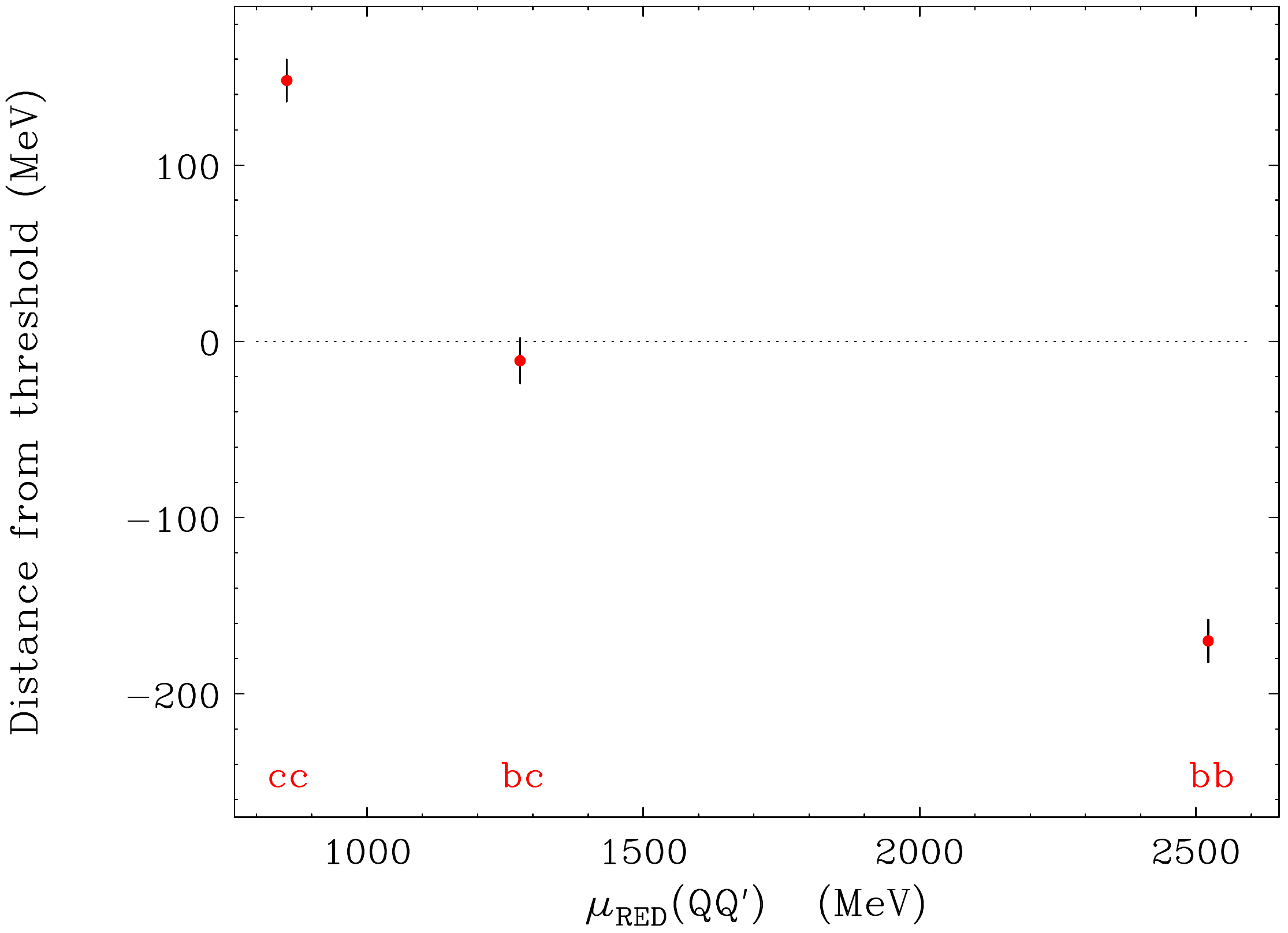}
\end{center}
\caption{\label{fig:threshold} 
Distance in MeV of the $cc\bar u\bar d$, $bc\bar u\bar d$ and $bb\bar u\bar d$
tetraquark masses from the corresponding thresholds $D^0 D^+ \gamma$,
$\bar B^0 D^0$, and $\bar B^0 B^- \gamma$, plotted against the reduced
masses of the doubly-heavy diquarks $\mu_{RED}(Q Q')$, $Q$,$Q'{=}c,b$.} 
\end{figure}

Fig.~\ref{fig:threshold} shows the distance in MeV between the masses of the 
$cc\bar u\bar d$, $bc\bar u\bar d$ and $bb\bar u\bar d$ tetraquarks
and the corresponding thresholds, $D^0 D^+ \gamma$, $\bar B^0 D^0$, and
$\bar B^0 B^- \gamma$, respectively, plotted against the reduced mass of the
doubly-heavy diquark. 

The main reason $bb\bar u\bar d$ is deeply bound, while $cc\bar u\bar d$ is
above threshold and $bc\bar u\bar d$ is borderline below
threshold, is the big jump in the $QQ$ binding energy as the heavy quarks'
mass increases: 129 MeV for $cc$ vs. $281$ MeV for $bb$.  This increase
in the binding energy can be understood qualitatively by noting that
the two heavy quarks are non-relativistic and their interaction can be
described by a Coulomb + linear potential, or by a logarithmic potential,
both of which are singular at the origin.  The mean distance between
the two heavy quarks scales like $1/(\alpha_s m_Q)$ and is significantly
smaller than the typical hadronic scale $\sim 1/\Lambda_{QCD}$.  At such
small distances as $m_Q$ increases, the $QQ$ binding energy grows rapidly
with shrinking distance, due to the singularity of the potential.

While the above provides a qualitative understanding of the phenomenon,
we stress again that the actual numerical value of the $QQ$ binding
energy employed here is not computed from any particular potential,
but rather taken directly from experiment, using the previously-discussed
correspondence between binding in $\bar QQ$ quarkonia and in $QQ$ diquarks
which led to the accurate prediction of the $\Xi_{cc}$ mass.

\section{$bb\bar u\bar d$ DECAY MODES AND LIFETIME \label{sec:decay}}
We focus on the decay of the $bb\bar u\bar d$ tetraquark which is
deeply bound, unlike $cc\bar u\bar d$ and $bc\bar u\bar d$ which are,
respectively, above and close to their relevant thresholds.

A crude estimate of the lifetime can be obtained similarly to
Ref.~\cite{Karliner:2014gca}.  We assume an initial state with mass
10,389.4 MeV, a final state with $M(\bar B)+M(D) = 7,144.5$ MeV, a charged weak
current giving rise to $e {\bar\nu}_e,\,\mu{\bar\nu}_\mu,\,\tau{\bar\nu}_\tau$
and three colors of $\bar u d$ and $\bar c s$, a kinematic suppression
factor
\beq
F(x) = 1 - 8x + 8 x^3 - x^4 + 12 x^2 \ln(1/x)~,
~~x \equiv \{[M(\bar B) + M(D)]/M(bb\bar u\bar d)\}^2,
\eeq
a value of $|V_{cb}| = 0.04$ as in Ref.\ \cite{Karliner:2014gca}, 
and a factor of 2 to count each decaying $b$ quark.  The resulting decay rate is
\beq
\Gamma(bb\bar u\bar d) = \frac{18~G_F^2 M(bb\bar u\bar d)^5}{192 \pi^3}
F (x) |V_{cb}|^2 = 17.9 \times 10^{-13}~{\rm
GeV}~,
\eeq
leading to a predicted lifetime $\tau(bb\bar u\bar d) = 367$ fs.
\medskip

\noindent
The $bb\bar u\bar d$ decay can occur through one of two channels: 
\medskip

\noindent
(a) The ``standard process" $bb \bar u \bar d \to c b \bar u \bar d + W^{*-}$.

\noindent
Typical reactions include\nl \vrule width 0pt height 2.5ex depth 1.5ex
    $T(bb\bar u\bar d) \to D^0 \bar B^0 \pi^-$,\, $D^+ B^- \pi^-$
\ and
\ $T(bb\bar u\bar d) \to \jpsi K^- \bar B^0$,\, $\jpsi \bar K^0 B^-$.

\noindent
In addition, there is a rare process where {\em both} $b$ quarks decay into
$c\bar c s$,
\nl
\vrule width 0pt height 2.5ex
   $T(bb\bar u\bar d) \to \jpsi \jpsi K^- \bar K^0$.
The signature for events with two $\jpsi$'s coming from the same secondary
vertex might be sufficiently striking to make it worthwhile to look for
such events against a large background.
\medskip

\noindent
(b) The $W$-exchange process $b \bar d \to c \bar u$, again involving either
one of the two $b$ quarks,
which ought to shorten the lifetime further.  The latter process can
involve a two-body final state, e.g., \,$T(bb\bar u\bar d) \to D^0 B^-$,\,
which may partially compensate for suppression due to
the small wave function of the $b \bar d$ pair at zero separation.
However, the comparable process in $B^0$ meson decay does not seem to shorten
its lifetime much with respect to $\tau(B^+)$.

\section{PRODUCTION \label{sec:prod}}

Production will be difficult because in addition to two $b$ quarks one will
need two $\bar b$ antiquarks.  The probability for producing two heavy quark
pairs can be estimated as the square of the probability for
producing one pair.  In the case of the doubly-charmed baryon $\Xi_{cc}^{++}$
observed by LHCb \cite{Aaij:2017ueg}, this difficulty appears to have been
overcome.

The signature for decay of a $b b \bar u \bar d$ state will be a final state
involving $b$ and $c$, whereas a $b \bar b$ state will give rise to a
$b$ and $\bar c$.  The mixing transition $D^0 \leftrightarrow \bar D^0$
in the latter process will induce a small background contribution to the
former process in final states containing a neutral $D$.  Similarly, final
states containing a $B^0$ will not be easily distinguishable from those
containing a $\bar B^0$ because of the mixing transition $B^0 \leftrightarrow
\bar B^0$.

Lipkin \cite{Lipkin:1986dw} has made the interesting point that in the limit of
very heavy $Q$, the $QQ\bar u \bar d$ color structure is that of an antibaryon.
On a related note, one can compare $QQ\bar u \bar d$ production with $QQq$
production achieved by LHCb in the discovery of $\Xi_{cc}^{++}$
\cite{Aaij:2017ueg}.  We assume that the relative fragmentation of the heavy
$QQ$ diquark into a light quark and into a light $(\bar u\bar d)_{I{=}0,S{=}0}$
diquark is analogous to the
relative fragmentation of a heavy antiquark (say, $\bar b$) into $u$ and into
$(\bar u \bar d)_{I{=}0,S{=}0}$.  These latter fragmentation fractions have
been measured for $b$ quarks produced at $\sqrt{s} = 1.96$ TeV at the
Fermilab Tevatron \cite{Aaltonen:2008zd}, with the results shown in Table
\ref{tab:frag}.  These imply central values $f_u = 0.356$, $f_d = 0.338$,
$f_s = 0.111$, and $f_{\Lambda_b} = 0.195$, assuming $f_u+f_d+f_s+f_{\Lambda_b}
=1$.  In other words, the fragmentation of a heavy $\bar Q$ into $\bar u\bar d$
in a state with $I=J=0$ occurs about half as frequently as fragmentation into
$u$.  
One can expect the same ratio for $bb \bar u \bar d$ relative to $bbu$.

\begin{table}
\caption{Fragmentation fractions of $b$ quarks produced at the Fermilab
Tevatron with $\sqrt{s} = 1.96$ TeV \cite{Aaltonen:2008zd}.  Errors are
statistical, systematic, and associated with branching fractions.
\label{tab:frag}}
\begin{center}
\strut\vskip-0.9cm\strut
\begin{tabular}{c c} \hline \hline
Ratio & Value \\ \hline
\vrule width 0pt height 3.0ex
      $f_u/f_d$      & $1.054 \pm 0.018^{+0.025}_{-0.045}\pm 0.058$ \\ 
\vrule width 0pt height 3.0ex
  $f_s/(f_u{+}f_d)$  & $0.160 \pm 0.005^{+0.011+0.057}_{-0.010-0.034}$ 
\kern1em\strut
\\
\vrule width 0pt height 3.0ex depth 1.5ex
$f_{\Lambda_b}/(f_u{+}f_d)$ & $0.281\pm0.012^{+0.058+0.128}_{-0.056-0.087}$ 
\kern1em\strut
\\
\hline \hline
\end{tabular}
\strut\vskip-1.0cm\strut
\end{center}
\end{table}

\section{SUMMARY\label{sec:summary}}

The calculation in Ref.\ \cite{Karliner:2014gca} makes use of a scheme in
which quarks in baryons are endowed with effective masses about 55 MeV heavier
than those in mesons.  An alternative approach with universal quark masses
compensates for this difference by adding a term $S = 165$ MeV associated with
a ``string junction,'' of which baryons possess one while mesons possess none
\cite{Karliner:2016zzc}.
Unless the $QQ$ system can be thought of as a pointlike object equivalent to
a heavy antiquark, a $Q Q \bar u \bar d$ tetraquark will have {\it two}
string junctions, with a corresponding increase in its mass.  An explicit
calculation shows that this increase is not enough to push the $b b \bar u
\bar d$ ground state mass above $B^+ B^0 \gamma$ threshold.

We have used the recent discovery by LHCb of a doubly-charmed baryon
$\Xi_{cc}^{++}$ \cite{Aaij:2017ueg} to confirm an assumption about the
interaction energy of two heavy quarks $Q$ in a tetraquark $T = QQ \bar u
\bar d$.  This has enabled us to estimate the mass of the $J^P = 1^+$ ground
state to be $10,389 \pm 12$ MeV, or 215 MeV below the threshold to decay
strongly to a $B B^*$ pair and 170 MeV below $B^- \bar B^0\gamma$ threshold.
Such a state will then decay only weakly,
initially via the subprocess $b \to c W^{*-}$.  The other $b$ eventually also
will decay via this subprocess, leaving a final state with two charmed quarks,
unless flavor-changing mixing of neutral $B$ or $D$ mesons has taken place.
The challenge will be to distinguish such a final state from one with $c
\bar c$.

Our approach is the first
to use the discovery of $\Xi_{cc}^{++}$ to ``calibrate'' the binding energy
in a $QQ$ diquark.  However, many other estimates exist of masses of
systems containing more than one heavy quark, by methods such as QCD sum rules,
potential models, heavy quark effective theory, and lattice gauge theory,
in addition to Ref.\ \cite{Luo:2017eub} and the many references cited therein.

An early paper to tackle such problems is Ref.~\cite{Ader:1981db}
which states in the Abstract ``[...] there is no [$QQ\bar Q \bar Q$]
state below the threshold corresponding to the spontaneous dissociation
into two mesons.'' Various potential ways out are discussed, including unequal
masses.  According to ref.~\cite{Lipkin:1986dw}, ``Four-quark states containing
two identical heavy quarks are shown to have a good probability of being
stable against strong decays.''  Recent treatments using QCD sum rules
\cite{Du:2012wp,Chen:2013aba} have an error of 200 MeV on a central
value $M(bb\bar u \bar d) = 10300$ MeV and therefore do not deliver a
crisp answer regarding stability of the $bb\bar u \bar d$ tetraquark
against strong decays.  An early estimate \cite{Janc:2003cf} concluded
that the $bb \bar u \bar d$ ground state tetraquark is stable against
strong and EM interactions, with a mass of $130 \pm 15$ MeV below $BB^*$
threshold and $85 \pm 15$ MeV below $B^0 B^+ \gamma$ threshold.
Ref.~\cite{Vijande:2007rf} finds that the ground state $c c \bar u \bar d$
tetraquark ``may be bound''; there is no discussion of the $bb\bar u
\bar d$ tetraquark.  This gives an idea of the spread of results.

Lattice QCD calculations have matured to the point that their calculated masses
of heavy-quark systems are stable to better than 10 or 20 MeV, and agree with
experiment to at least that accuracy.  An encouraging result
\cite{Francis:2016hui} finds the lowest $bb\bar u \bar d$ state $144 \pm 10$
MeV below $B^- \bar B^0\gamma$ threshold, not far from our value of $170 \pm
12$ MeV.
The lowest $bb \bar s \bar q$ state ($q=u$ or $d$) is found to be $52 \pm 8$
MeV below $B_s B \gamma$ threshold.

Experimental search for the $bb\bar u\bar d$ tetraquark is a challenge well
worth pursuing, because it is the first manifestly exotic hadron stable under
strong and EM interactions.

\section*{ACKNOWLEDGMENTS}

We thank V. Belyaev, S.~Eidelman, R.~Lewis, and T. Skwarnicki for comments on
the manuscript, J.-M. Richard for directing us to some relevant literature, and
all our colleagues in the LHCb Exotics Group for encouragement.  The work of
J.L.R.  was supported in part by the U.S. Department of Energy, Division of
High Energy Physics, Grant No.\ DE-FG02-13ER41958.

\end{document}